\documentclass[11pt]{article}

\makeatletter
\@addtoreset{equation}{section}
\makeatother
\usepackage{amsmath}
\usepackage{amsthm}
\usepackage{amsfonts}
\usepackage{amssymb}
\usepackage{bm}
\usepackage{graphicx} 
\usepackage{subfigure} 
\usepackage{float} 
\usepackage{color}
\usepackage{mathrsfs}
\usepackage{euscript}
\usepackage{fancyheadings}

\theoremstyle{remark}

\newcommand{\cA}{\mathcal{A}}

\newcommand{\cC}{\mathcal{C}}

\newcommand{\cH}{\mathcal{H}}
\newcommand{\cL}{\mathcal{L}}

\newcommand{\cI}{\mathcal{I}}
\newcommand{\cR}{\mathcal{R}}

\newcommand{\cM}{\mathcal{M}}

\newcommand{\bZ}{\mathbb{Z}}

\newcommand{\be}{\begin{equation}}
\newcommand{\ee}{\end{equation}}
\newcommand{\bea}{\begin{eqnarray}}
\newcommand{\eea}{\end{eqnarray}}
\newcommand{\non}{\nonumber}

\begin{document}

\begin{center}
\begin{Large}
{\bf Boundary bound states in the SUSY sine-Gordon model with Dirichlet boundary conditions}
\end{Large}

\vspace{65pt}

\begin{large}
{\bf Chihiro Matsui}$^{1,2}$\footnote[5]{matsui@sat.t.u-tokyo.ac.jp}
\end{large}
\par
\vspace{20pt}
\parbox{11cm}{\it 
$^1$ FIRST, Aihara Innovative Mathematical Modelling Project, Japan Science and Technology Agency, 4-6-1, Komaba, Meguro-ku, Tokyo, 153-8505, Japan

$^2$ Institute of Industrial Science, The University of Tokyo, 4-6-1, Komaba, Meguro-ku, Tokyo, 153-8505, Japan
}
\end{center}

\vspace{40pt}

\begin{center}
\centerline{\bf \small Abstract} \vspace{20pt}
\parbox{11cm}
{\small 
We analyze the ground state structure of the supersymmetric sine-Gordon model via the lattice regularization. 
The nonlinear integral equations are derived for any values of the boundary parameters by the analytic continuation and showed three different forms depending on the boundary parameters. We discuss the state that each set of the nonlinear integral equations characterizes in the absence of source terms. 
Four different pictures of the ground state are found by numerically studying the positions of zeros in the auxiliary functions. We suggest the existence of two classes in the SUSY sine-Gordon model, which cannot be mixed each other.
}
\end{center}

\vspace{40pt}

\section{Introduction}
Boundary systems have provided many interesting phenomena including the edge states and the boundary critical exponents. 
Also as any real materials are finite systems, it is important to know how the boundaries affect on physical quantities. Nevertheless, analytical derivation of physical quantities often becomes drastically difficult due to existence of boundaries. 

Integrable systems give possibility for analytic calculation of boundary systems. Some examples are known to keep the system integrable even after imposing non-periodic boundary conditions. The famous one is the Heisenberg spin chain with boundary magnetic fields. This model was solved through the $q$-deformed vertex operator method \cite{bib:JKKKM95} and the Bethe ansatz method \cite{bib:SS95,bib:KS96}. The interesting feature was found that the ground state is encoded by different structures of the roots depending on strength of boundary magnetic fields. 

The XXZ model with boundary magnetic fields appears also in the context of the integrable quantum field theory through the lattice regularization. The sine-Gordon model with the Dirichlet boundary conditions is a famous example, whose lattice regularized model is known as the XXZ model with boundary magnetic fields. 
As an integrable system, the $S$-matrix of the sine-Gordon model satisfies relations called the Yang-Baxter equation, which allows us to know the exact expressions of the scattering amplitudes. Also, the reflection relation is known for the integrable open system, from which the exact reflection amplitudes can be obtained. There are some particles which do not scatter of each other but form bound states. For those cases, the procedure called the bootstrap principle has been developed to calculate the scattering amplitudes between bound states and their mass starting from the soliton mass, i.e. the mass of the lightest particle. 

However, few analytic results are available for non-periodic systems, especially for systems with complicated symmetries such as supersymmetry. 
%
For this reason, we study boundary effect on the ground state of the supersymmetric (SUSY) sine-Gordon model with the Dirichlet boundary conditions via the lattice regularized model in this paper. 
Actually, knowing the ground state structure has the meanings more than just because the ground state is the most basic object. As was obtained for the spin-$1/2$ case, the configuration of the ground state roots does affect on the multiple-integral expressions of correlation functions \cite{bib:KKMNST07, bib:KKMNST08}. 
From the view point of the inhomogeneous XXZ model, the SUSY sine-Gordon model is the integrable extension to the spin-$1$ model. On the other hand, in the quantum field theoretic viewpoint, the SUSY sine-Gordon model is understood as the quantum field theory described by solitons characterized not only by the soliton charge but also by the RSOS indices. 
Unlike the original sine-Gordon case, few results are available for the SUSY case. This is partly because of complicated structure of the ground state, which depends on boundary parameters. 

Our main aim is studying the ground state structure of the SUSY sine-Gordon model with the Dirichlet boundary conditions for any value of boundary parameters. To accomplish this aim, we first derive the nonlinear integral equations based on the method introduced in \cite{bib:KBP91}, from which one can read off the reflection amplitude. 
Some useful facts about the SUSY sine-Gordon model are reviewed in Section \ref{sec:SUSYSG}. In Section \ref{sec:NLIE}, the nonlinear integral equations are derived. Performing the analytic continuation of them with respect to boundary parameters, we obtain three different forms each of which describe scatterings on different states. After giving explanation about the states each set of the nonlinear integral equations is describing, 
%
the correct ground state is determined for full range of boundary parameters in Section \ref{sec:bbs}. 
The numerical plots of zeros in the auxiliary functions are given and the structure of the ground state is discussed from the analysis of those zeros. 
In the last section, we give concluding remarks and open problems.

\section{Supersymmetric sine-Gordon model with Dirichlet boundary conditions}\label{sec:SUSYSG}
We first give a brief review of the SUSY sine-Gordon model. 
The SUSY sine-Gordon model is a quantum field theory described by the following action: 
\begin{equation}\label{SSG_action}
\begin{split}
&\cA_{\rm SSG} = \int_{-\infty}^{\infty} dt \int_{-\infty}^{\infty} dx\; \cL_{\rm SSG}(x,t)
\\
&\cL_{\rm SSG} = \frac{1}{2} \partial_{\mu} \varphi \partial^{\mu} \varphi + \frac{i}{2} \bar{\psi} \gamma^{\mu} \partial_{\mu} \psi - \frac{\cM}{2} \cos(\beta \varphi) \bar{\psi} \psi + \frac{\cM^2}{2\beta^2} \cos^2(\beta \varphi), 
\end{split}
\end{equation}
which consists of a real scalar feld $\varphi$ and a Majorana fermion field $\psi$. A soliton mass is denoted by $\cM$ and a dimensionless coupling constant by $\beta$. The value of $\beta$ determines a main feature of the theory; the regime $0<\beta^2<4\pi/3$ is known as an attractive regime and $4\pi/3<\beta^2<4\pi$ a repulsive regime. 

Though the action (\ref{SSG_action}) is defined on an infinite length, it is also possible to consider the theory on a finite length:  
\begin{equation}\label{BSSG_action}
\cA_{\rm BSSG}=\int_{-\infty}^{\infty}dt\int_{0}^L dx\;\cL_{\rm SSG}(x,t). 
\end{equation}
Special forms of boundary conditions are required in order to keep integrability of the system \cite{bib:S88} and here we choose \emph{the Dirichlet boundary conditions}: 
\begin{equation}
\begin{split}
&\varphi(x,t)\Big|_{x=0}=\varphi_-\qquad
\psi(x,t)-\bar{\psi}(x,t)\Big|_{x=0}=0\\
&\varphi(x,t)\Big|_{x=L}=\varphi_+\qquad
\psi(x,t)-\bar{\psi}(x,t)\Big|_{x=L}=0.
\end{split}
\end{equation}
Let us remark that there is another set of the Dirichlet boundary conditions, although we will not consider it hereafter, given by 
\begin{equation}
\psi(x,t)+\bar{\psi}(x,t)\Big|_{x=0}=0
\qquad
\psi(x,t)+\bar{\psi}(x,t)\Big|_{x=L}=0. 
\end{equation}

\subsection{$S$-matrix theory}
One of the main virtues to work on integrable systems is one can know the exact $S$-matrix by solving the characteristic relations satisfied by the $S$-matrix. The $S$-matrix of the quantum field theory is defined for the asymptotic states: 
\begin{equation}
\begin{split}
&|A_{a_0 a_1}^{\sigma_1}(\theta_1) A_{a_1 a_2}^{\sigma_2}(\theta_2)\dots A_{a_{M-1} a_M}^{\sigma_M}(\theta_M)\rangle
\\
&=A_{a_0 a_1}^{\sigma_1}(\theta_1) A_{a_1 a_2}^{\sigma_2}(\theta_2)\dots A_{a_{M-1} a_M}^{\sigma_M}(\theta_M)|0\rangle 
\end{split}
\end{equation}
created by applying the operators $A_{a_{j-1} a_{j}}^{\sigma_j}(\theta_j)$ to the vacuum $|0\rangle$. Rapidity of each particle $\theta_j$ obeys the condition $\theta_1>\dots>\theta_M$ for an initial state and $\theta_1<\dots < \theta_M$ for a final state. The soliton creation operator has two types of indices one of which denotes the soliton charge $\sigma_j=\pm$ and the other expresses the RSOS indices $a_j=0,\frac{1}{2},1$ under the condition $|a_j - a_{j-1}| = \frac{1}{2}$. 

The $S$-matrix $S_{\rm SSG}(\theta)$ determines scattering amplitudes of a two-particle scattering process: 
\begin{equation}
A_{ab}^{\sigma_1}(\theta_1)+A_{bc}^{\sigma_2}(\theta_2)\rightarrow 
A_{ad}^{\sigma'_2}(\theta_2)+A_{dc}^{\sigma'_1}(\theta_1). 
\end{equation}
From the known properties of the $S$-matrix: the initial condition, the unitarity condition, the crossing symmetry, and the Yang-Baxter equation originated from integrability of the theory together with decomposability of the SUSY sine-Gordon $S$-matrix: 
\begin{equation}
S_{\rm SSG}(\theta) = S_{\rm SG}(\theta) \otimes S_{\rm RSOS}(\theta), 
\end{equation}
the exact expressions of scattering amplitudes are obtained. 
%
%
We skip the derivation of the $S$-matrix, as it was closely discussed in the literatures \cite{bib:Z79, bib:A91, bib:GZ94, bib:AK96, bib:NA02} and just show the final results. The sine-Gordon part is given by
\begin{equation}
S_{\rm SG} (\theta) = 
\begin{pmatrix}
S(\theta) & 0 & 0 & 0 \\
0 & S_T(\theta) & S_R(\theta) & 0 \\
0 & S_R(\theta) & S_T(\theta) & 0 \\
0 & 0 & 0 & S(\theta)
\end{pmatrix}.
\end{equation}
The transmissive and reflective elements of the $S$-matrix $S_T(\theta)$ and $S_R(\theta)$ are expressed by the soliton-soliton scattering amplitude $S(\theta)$ through the relation: 
\begin{equation}
\begin{split}
&S_T(\theta)=\frac{\sinh\lambda \theta}{\sinh\lambda(i\pi-\theta)}S(\theta)
\\
&S_R(\theta)=i\frac{\sin\pi\lambda}{\sinh\lambda(i\pi-\theta)}S(\theta)
\end{split}
\end{equation}
which, by using the notation $u:= -i\theta$, is given by
\begin{equation}
\begin{split}
&S(\theta)=-\prod_{l=1}^{\infty} \left[
\frac{\Gamma(2(l-1)\lambda - \frac{\lambda u}{\pi}) \Gamma (2l\lambda +1 -\frac{\lambda u}{\pi})}
{\Gamma ((2l-1)\lambda - \frac{\lambda u}{\pi}) \Gamma ((2l-1)\lambda + 1 - \frac{\lambda u}{\pi})}/(u\rightarrow -u)
\right]. 
\end{split}
\end{equation}
On the other hand, the RSOS part is written as 
\begin{equation}
\begin{split}
&S_{\rm RSOS}\left(
\left.
\begin{matrix}
a & d \\ b & c
\end{matrix}\,
\right|
\theta
\right)
= X_{ad}^{bc}(\theta) K(\theta)
\\
&K(\theta) = \frac{1}{\sqrt{\pi}}
\prod_{k-1}^{\infty} \frac{\Gamma (k-\frac{1}{2}+\frac{\theta}{2\pi i}) \Gamma (k-\frac{\theta}{2\pi i})}
{\Gamma(k+\frac{1}{2}-\frac{\theta}{2\pi i}) \Gamma (k+\frac{\theta}{2\pi i})}. 
\end{split}
\end{equation}
The function $X_{ad}^{bc} (\theta)$ is given for each set of the RSOS indices: 
\begin{equation}
\begin{split}
&X_{\frac{1}{2} 0}^{0 \frac{1}{2}} (\theta) = X_{\frac{1}{2} 1}^{1 \frac{1}{2}} (\theta) = 2^{(i\pi - \theta)/2\pi i} \cos\left(\frac{\theta}{4i} - \frac{\pi}{4}\right)
\\
&X_{0 \frac{1}{2}}^{\frac{1}{2} 0} (\theta) = X_{1 \frac{1}{2}}^{\frac{1}{2} 1} (\theta) = 2^{\theta/2\pi i} \cos\left(\frac{\theta}{4i}\right)
\\
&X_{\frac{1}{2} 1}^{0 \frac{1}{2}} (\theta) = X_{\frac{1}{2} 0}^{1 \frac{1}{2}} (\theta) = 2^{(i\pi - \theta)/2\pi i} \cos\left(\frac{\theta}{4i} + \frac{\pi}{4}\right)
\\
&X_{0 \frac{1}{2}}^{\frac{1}{2} 1} (\theta) = X_{1 \frac{1}{2}}^{\frac{1}{2} 0} (\theta) = 2^{\theta/2\pi i} \cos\left(\frac{\theta}{4i} - \frac{\pi}{2}\right). 
\end{split}
\end{equation}

Moreover, for the finite-size theory, scatterings on boundaries are ruled by \emph{the reflection matrix}: 
\begin{equation}
R_{\rm SSG}(\theta) = R_{\rm SG}(\theta) \otimes R_{\rm RSOS}(\theta). 
\end{equation}
The exact expression of the reflection matrix is obtained from the reflection relations for integrable boundaries. The sine-Gordon part of the reflection matrix for the Dirichlet boundary conditions is given by a $2$-by-$2$ matrix: 
\begin{equation}
\begin{split}
&R_{\rm SG}(\theta)=
\begin{pmatrix}
\cos(\xi+\lambda u) & 0 \\
0 & \cos(\xi-\lambda u)
\end{pmatrix}
R_0(u) \frac{\sigma(\theta, \xi)}{\cos \xi}
\\
&R_0(u)=
\prod_{l=1}^{\infty} \left[
\frac{\Gamma(4l\lambda - \frac{2\lambda u}{\pi}) \Gamma (4\lambda (l-1) + 1 -\frac{2\lambda u}{\pi})}
{\Gamma(4l-3) \lambda - \frac{2\lambda u}{\pi} \Gamma((4l-1)\lambda +1 -\frac{2\lambda u}{\pi})}
/ (u\rightarrow -u)
\right]
\\
&\sigma(\theta, \xi) = \frac{\cos\xi}{\cos(\xi+\lambda u)}
\\
&\times
\prod_{l=1}^{\infty}\left[
\frac{\Gamma(\frac{1}{2} + \frac{\xi}{pi} + (2l-1)\lambda -\frac{\lambda u}{\pi}) \Gamma (\frac{1}{2} - \frac{\xi}{\pi} + (2l-1)\lambda -\frac{\lambda u}{\pi})}
{\Gamma(\frac{1}{2} - \frac{\xi}{\pi} + (2l-2)\lambda - \frac{\lambda u}{\pi}) \Gamma(\frac{1}{2} + \frac{\xi}{\pi} + 2l\lambda - \frac{\lambda u}{\pi})}
/(u\rightarrow -u)
\right],
\end{split}
\end{equation}
while the RSOS part is obtained as
\begin{equation}
\begin{split}
& R_{\frac{1}{2} \frac{1}{2}}^0(\theta) R_{\frac{1}{2} \frac{1}{2}}^1(\theta)
= 2^{-\theta/\pi i} P(\theta)
\\
& P(\theta) = \prod_{k=1}^{\infty} \left[
\frac{\Gamma (k - \frac{\theta}{2\pi i}) \Gamma(k - \frac{\theta}{2\pi i})}
{\Gamma(k-\frac{1}{4} - \frac{\theta}{2\pi i}) \Gamma(k+\frac{1}{4} - \frac{\theta}{2\pi i})}
/(\theta \rightarrow -\theta)
\right]. 
\end{split}
\end{equation}
Here we gave the expression of the RSOS reflection matrix only for $R_{\frac{1}{2} \frac{1}{2}}^a(\theta)$. As the ground state is non-degenerate, we assume the RSOS index for the ground state is $\frac{1}{2}$. Therefore, the other RSOS reflection factors are expected to be obtained for the reflection on an excited boundary. For this reason, the expression for $R_{ab}^{\frac{1}{2}}(\theta)$ ($a,b=0,1$) will be given in the context of the boundary bootstrap approach in the next subsection.

\subsection{Bound states}
Particles may form bound states in the process of many-body scatterings. Bound states are associated with poles in the $S$-matrix. The bootstrap approach has been developed, which allows us to know the scattering amplitudes between bound states and their masses. 
The same strategy is available to boundary reflections, which is called \emph{the boundary bootstrap}. 
Poles of the reflection matrix give rapidity of boundary bound states: 
\begin{equation}
\begin{split}
&\nu_n = \frac{\xi}{\lambda} - \frac{\pi (2n + 1)}{2\lambda}
\\
&w_N  = \pi - \frac{\xi}{\lambda} - \frac{\pi (2N-1)}{2\lambda}. 
\end{split}
\end{equation}
Only those poles which are in the physical strip $\cI m\;\theta\in (0,\frac{\pi}{2})$ contribute to boundary excitations. The condition to obtain at least one boundary bound state is then derived as $\xi > \pi/2$. Thus, particles trapped at the boundary appear when the boundary parameter exceed the threshold. 
Let us make a remark on the restriction for $n$ and $N$. Through this paper, we restrict out interest to the repulsive regime $0<\lambda<1$. Therefore, when $\theta = i\nu_0$ gets into the physical strip, the others are outside and thus we obtain only one boundary bound state.

The boundary bound states form excited boundaries on which the reflection amplitudes are read off from the boundary bootstrap. The bootstrap principle for the RSOS reflection factor is given by
\begin{equation}
g_{|a,\frac{1}{2}|\nu_n\rangle}^{|\frac{1}{2}\rangle} R_{ab}^{\frac{1}{2}} (\theta) = g_{|b,\frac{1}{2}|\nu_n\rangle}^{|\frac{1}{2}\rangle} \left[
\sum_{x=0,1} 
S\begin{pmatrix}\frac{1}{2} & x \\ a & \frac{1}{2} \end{pmatrix} (\theta - i\nu_n)
S\begin{pmatrix}\frac{1}{2} & b \\ x & \frac{1}{2} \end{pmatrix} (\theta + i\nu_n)
R_{\frac{1}{2} \frac{1}{2}}^x(\theta)
\right], 
\end{equation}
which leads to the RSOS reflection amplitude on an excited boundary: 
\begin{equation}
R_{ab}^{\frac{1}{2}} (\theta) = P(\theta) K(\theta-i\nu_n) K(\theta+i\nu_n)
\frac{g_{|b,\frac{1}{2}|\nu_n\rangle}^{|\frac{1}{2}\rangle}}{g_{|a,\frac{1}{2}|\nu_n\rangle}^{|\frac{1}{2}\rangle}}
\left(\delta_{ab}\cos\frac{\nu_n}{2} + \delta_{a,1-b}\sin\frac{\theta}{2i}\right),  
\end{equation}
where the $g$-factor is the SUSY part of the boundary coupling \cite{bib:LMSS95,bib:BPT03}.

\subsection{Lattice regularization}
 The discretization of quantum field theories is called \emph{the lattice regularization} and found for some models including the sine-Gordon model. Such a notion was first introduced by Destri and de Vega \cite{bib:DV87}. 
For instance, the sine-Gordon model can be lattice-regularized into the inhomogeneous spin-$1/2$ XXZ model, or inversely, one can say that the spin-$1/2$ XXZ model has the sine-Gordon model as its low-energy effective field theory. 
These two models are equivalent in the scaling limit, where the inhomogeneity $\Theta \rightarrow \infty$, the site numbers $N\rightarrow \infty$, and the lattice spacing $a\rightarrow 0$ by keeping the soliton mass $\cM:= 4e^{-\Theta}/a$ finite \cite{bib:RS94}. 
In the same way, the SUSY sine-Gordon model is also discretized into a lattice system. The corresponding model is the inhomogeneous spin-$1$ XXZ chain as was verified in \cite{bib:IO92}
\begin{figure}[!h]
\begin{center}
\includegraphics[scale=0.45]{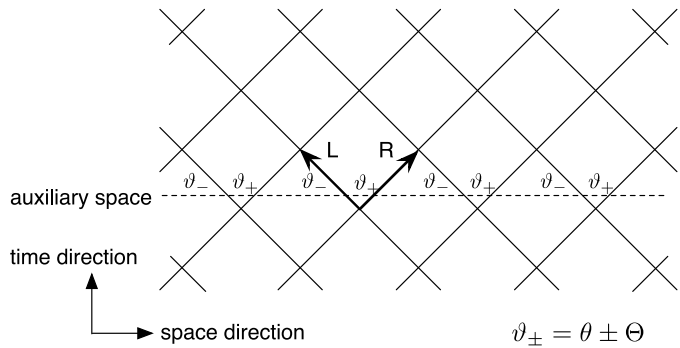}
\caption{The light-cone lattice. The inhomogeneity $\pm\Theta$ is assigned respectively to a right-moving particle (R) and a left-moving particle (L). }
\end{center}
\end{figure}

Lattice regularization provided another way to analyze the continuum theory; though the Bethe ansatz method can be applied only to discretized models, the lattice sine-Gordon model admits this treatment. As relativistic field theory, the sine-Gordon model can be considered as a theory defined on the light cone, which is regarded as a Minkowski space-time. The space-time development on the light cone can be described by the $R$-matrices with inhomogeneity $\pm \Theta$ as rapidity of a right-moving particle (or, respectively, that of a left-moving particle). We skip the details which can be found in \cite{bib:DV87}.

\section{Nonlinear integral equations}\label{sec:NLIE}
The scattering process on the lattice was understood through the counting function in the spin-$1/2$ case. The counting function first derived by Destri and de Vega \cite{bib:DV92,bib:DV95} gives a \emph{counting} of the roots which contribute to the ground state, i.e. the real roots for the spin-$1/2$ case. With the use of the $T$-function, the object related to the transfer matrix, one can recast the thermodynamic Bethe ansatz equations. The sub-leading terms in the thermodynamic Bethe ansatz equations lead to the scattering amplitudes. It is known that all the elements of the scattering matrix can be obtained through the classification of the Bethe roots.

Unlikely, the definition of the counting function is unclear for the spin-$1$ case, due to the $2$-string structure of the ground state. Instead, substitution has been developed in \cite{bib:ANS07}. The method is based on the auxiliary functions \cite{bib:KBP91}, which is defined through two $T$-functions coming from two valid transfer matrices $T_1(\theta)$ and $T_2(\theta)$ of the spin-$1$ XXZ model.


The scattering processes can be read of from the nonlinear integral equations derived based on the analytic and non-zero property of the $T$-functions. The sub-leading terms correspond to particle-scattering and boundary-scattering amplitudes. The source terms in the nonlinear integral equations are understood as the effect of excitation particles. A soliton is obtained as an ordinary hole in the distribution of the two-strings. The interpretation of some other types of particles can be found in the literatures \cite{bib:DV97,bib:FRT00}

Such nonlinear integral equations were closely studies in \cite{bib:S04, bib:HRS06, bib:H07} under the periodic boundary condition. In this section, we derive the nonlinear integral equations with the Dirichlet boundaries.  

\subsection{Auxiliary functions}
\begin{figure}[!h]
\begin{center}
\includegraphics[scale=0.5]{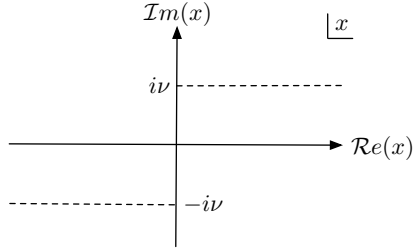}
\caption{The branch cut of the logarithm.}\label{fig:branch}
\end{center}
\end{figure}

The auxiliary functions are related to two relevant transfer matrices of the spin-$1$ XXZ model, $T_1(\theta)$ and $T_2(\theta)$, which are explicitly written as
\begin{equation}
\begin{split}
&T_1(\theta)=l_1(\theta)+l_2(\theta)\\
&T_2(\theta)=\lambda_1(\theta)+\lambda_2(\theta)+\lambda_3(\theta).
\end{split}
\end{equation}
Each function in $T_1(\theta)$ and $T_2(\theta)$ is given by
\begin{equation}
\begin{split}
&l_1(\theta)=\sinh(2\theta+i\pi)B_+(\theta)\phi(\theta+i\pi)\frac{Q(\theta-i\pi)}{Q(\theta)}\\
&l_2(\theta)=\sinh(2\theta-i\pi)B_-(\theta)\phi(\theta-i\pi)\frac{Q(\theta+i\pi)}{Q(\theta)}
\end{split}
\end{equation}
and 
\begin{equation}
\begin{split}
&\lambda_1(\theta)=\sinh(2\theta-2i\pi)B_-\left(\theta-\frac{i\pi}{2}\right)B_-\left(\theta+\frac{i\pi}{2}\right)
\\
&\hspace{14mm}
\times\phi\left(\theta-\frac{3i\pi}{2}\right)\phi\left(\theta-\frac{i\pi}{2}\right)\frac{Q(\theta+\frac{3i\pi}{2})}{Q(\theta-\frac{i\pi}{2})}
\\
&\lambda_2(\theta)=\sinh(2\theta)B_+\left(\theta-\frac{i\pi}{2}\right)B_-\left(\theta+\frac{i\pi}{2}\right)
\\
&\hspace{14mm}
\times\phi\left(\theta-\frac{i\pi}{2}\right)\phi\left(\theta+\frac{i\pi}{2}\right)\frac{Q(\theta+\frac{3i\pi}{2})Q(\theta-\frac{3i\pi}{2})}{Q(\theta-\frac{i\pi}{2})Q(\theta+\frac{i\pi}{2})}
\\
&\lambda_3(\theta)=\sinh(2\theta+2i\pi)B_+\left(\theta-\frac{i\pi}{2}\right)B_+\left(\theta+\frac{i\pi}{2}\right)
\\
&\hspace{14mm}
\times\phi\left(\theta+\frac{3i\pi}{2}\right)\phi\left(\theta+\frac{i\pi}{2}\right)\frac{Q(\theta-\frac{3i\pi}{2})}{Q(\theta+\frac{i\pi}{2})},
\end{split}
\end{equation}
where
\begin{equation}
\begin{split}
&\phi(\theta)=\sinh^M(\theta-\Theta)\sinh^M(\theta+\Theta)\\
&B_{\pm}(\theta)=\sinh\left(\theta\pm\frac{i\pi H_+}{2}\right)\sinh\left(\theta\pm\frac{i\pi H_-}{2}\right)\\
&Q(\theta)=\prod_{k=1}^m\sinh(\theta-\theta_k)\sinh(\theta+\theta_k). 
\end{split}
\end{equation}
The variables $\theta_k$ are taken as Bethe roots with positive real parts. Thus, the function $Q(\theta)$ gives roots as its zeros. 
The ambiguity of $2\pi i$-periodicity shall be removed by choosing the branch of a function: 
\begin{equation}
f_{\nu}(x)=\frac{1}{i}\ln\left[-\frac{\sinh(x-i\nu)}{\sinh(x+i\nu)}\right]
\end{equation}
as in Figure \ref{fig:branch}.

We define the auxiliary functions as
\begin{align}
&b(\theta)=\frac{\lambda_1(\theta)+\lambda_2(\theta)}{\lambda_3(\theta)}
\qquad
\bar{b}(\theta)=\frac{\lambda_3(\theta)+\lambda_2(\theta)}{\lambda_1(\theta)}=b(-\theta)
\\ 
&y(\theta)=\frac{T_0(\theta)T_2(\theta)}{f(\theta)}
\\
&B(\theta)=1+b(\theta)
\qquad
\bar{B}(\theta)=1+\bar{b}(\theta)
\qquad
Y(\theta)=1+y(\theta), 
\end{align}
where
\begin{align}
&T_0(\theta)=\sinh(2\theta)\\
&f(\theta)=l_2\left(\theta-\frac{i\pi}{2}\right)l_1\left(\theta+\frac{i\pi}{2}\right).
\end{align}
It is also useful to define the ratio of $l_1(\theta)$ and $l_2(\theta)$ by
\begin{equation}
a_1(\theta) := \frac{l_2(\theta)}{l_1(\theta)}, 
\end{equation}
which is related to $T_1(\theta)$ and $b(\theta)$ via
\begin{align}
& T_1(\theta) = l_2(\theta) (1 + a_1^{-1}(\theta)) 
\\
& b(\theta) = a_1\left(\theta+\frac{i\pi}{2}\right) a_1\left(\theta-\frac{i\pi}{2}\right) \left(1 + a_1^{-1}\left(\theta - \frac{i\pi}{2}\right)\right). 
\end{align}
Holes in distribution of the two-strings are obtained as zeros in the transfer matrix $T_2(\theta)$ and those in $T_1(\theta)$ are understood as holes in real roots. 
On the other hand, the auxiliary function $B(\theta)$ has zero, besides as holes, at two-string centers, and similarly, $1+a_1(\theta)$ at roots themselves.

From these facts, the meanings of the auxiliary functions $b(\theta)$ and $a_1(\theta)$ are understood as follows; the function $b(\theta)$ gives a counting of the number of two-string pairs and $a_1(\theta)$ a counting of real roots. Indeed, the logarithms of the two auxiliary functions  count these quantities, being associated with the quantum numbers $I_j \in \bZ+\frac{1}{2}$ \cite{bib:H07}: 
\begin{equation}
i \ln a_1(\theta_j) = 2\pi I_j
\qquad 
i \ln b\left(\theta_i-\frac{i\pi}{2}\right) = 2\pi I_j. 
\end{equation}
As the ground state is encoded by the two-string roots, real roots should be considered as excited particles. For this reason, we name hole-type solutions of $1+a_1(\theta)=0$ as type-1 holes.

The transfer matrix $T_2(\theta)$ can be written by the auxiliary functions in two different ways, which we will use later to deform the contour of integrals: 
\begin{align}
T_2(\theta)&=t_+(\theta)\frac{Q(\theta-\frac{3i\pi}{2})}{Q(\theta+\frac{i\pi}{2})} B(\theta)
\\
&=t_-(\theta)\frac{Q(\theta+\frac{3i\pi}{2})}{Q(\theta-\frac{i\pi}{2})} \bar{B}(\theta),
\end{align}
where
\begin{equation}
t_{\pm}(\theta)=\sinh(2\theta\pm 2i\pi)
B_{\pm}\left(\theta-\frac{i\pi}{2}\right) B_{\pm}\left(\theta+\frac{i\pi}{2}\right)
\phi\left(\theta\pm\frac{3i\pi}{2}\right) \phi\left(\theta\pm\frac{i\pi}{2}\right). 
\end{equation}
The transfer matrix $T_1(\theta)$ is expressed by another auxiliary function from the fusion relation:
\begin{equation}
T_1\left(\theta-\frac{i\pi}{2}\right) T_1\left(\theta+\frac{i\pi}{2}\right)
=f(\theta) Y(\theta). 
\end{equation}

\subsection{Classification of roots and holes}
The ground state structure of the XXZ spin chain reveals strong dependence on the anisotropy. For instance, when $\cos\gamma > 1$ the system is massive, while massless when $\cos \gamma \leq 1$. Here we focus on the case with $t:=\pi/\gamma\in\bZ_{\ge 3}$, where the ground state is written exactly by the two-strings in the thermodynamic limit of the periodic system. 
In the context of the SUSY sine-Gordon model, this parameter range falls into the repulsive regime $0<\lambda:=(t-2)^{-1}<1$, in which no breathers are allowed to exist. 

Bethe roots appear either as real numbers or as complex conjugate pairs except for the self-conjugate roots. It is useful to classify the roots into the four types depending on the values of their imaginary parts: 
\begin{itemize}
\item Inner roots $x_j$ ($j=1,2,\dots, M_I$)
	\hspace{21mm} $|\cI m(x_j)|<\frac{\pi}{2}$
\item Close roots $c_j$ ($j=1,2,\dots,M_C$)
	\hspace{20mm} $\frac{\pi}{2}<|\cI m(c_j)|<\frac{3\pi}{2}$
\item Wide roots $w_j$ ($j=1,2,\dots,M_W$)
	\hspace{18mm} $\frac{3\pi}{2}<|\cI m(w_j)|<\frac{\pi^2}{2\gamma}$
\item Self-conjugate roots $w^{SC}_{j}$ ($j=1,2,\dots,M_{SC}$)	
	\, $|\cI m(w^{SC}_j)|=\frac{\pi^2}{2\gamma}$. 
\end{itemize}
Let us remark that pure imaginary roots or holes, which belong to neither of these four regimes, appear due to the Dirichlet boundaries. However, here we first consider the regime $H_{\pm}>1$ where the ground state is given by the pure two-string roots in spite of the Dirichlet boundaries. The exact two-string structure of the ground state is verified also from the numerical results \cite{bib:AN04, bib:M11}.  
The reason to choose this regime is partly because we cannot help relying on the numerical results to know the correct ground state in the other regimes, in which boundary bound states may appear. Mostly, we show the similar statement given for the spin-$1/2$ case \cite{bib:ABR04} that the analytic continuation in the Fourier transform of the nonlinear integral equations with respect to the boundary parameters $H_{\pm}$ naturally leads to those with pure imaginary objects. 

We write holes in the two valid $T$-matrices of the SUSY sine-Gordon model by $h_j^{(1)}$ and $h_j$. Since the $T$-matrices have zeros at the positions of holes, we have the following relations: 
\begin{equation}
\begin{split}
&T_1(h^{(1)}_j)=0 \qquad j=1,2,\dots,N_{1}
\\
&T_2(h_j)=0 \hspace{10mm} j=1,2,\dots, N_H, 
\end{split}
\end{equation}
where the number of holes and type-1 holes are given by $N_1$ and $N_H$, respectively. Here we employ the assumption that both types of holes lie on the real axis, as in the periodic boundary system \cite{bib:S04, bib:HRS06, bib:H07}.

There also exist special types of holes and roots, on which $B(\theta)$ crosses the cut of the logarithm. The imaginary part of the logarithm of $b(\theta)$ has the derivative of a negative value:  
\begin{align*}
&i \ln b(s_j) = 2\pi I_{s_j}\qquad |b(s_j)| > 1\qquad (i \ln b)'(s_j) < 0\qquad j=1,\dots, N_S. 
\end{align*}
Then the logarithm of the auxiliary function $B(\theta)$ should be modified as 
\begin{equation}
\ln B(\theta) \rightarrow \ln B(\theta) 
+ 2\pi i \sum_{j=-N_S}^{N_S} H(\theta-s_j), 
\end{equation}
where $H(\theta)$ is the Heviside step function. 

\subsection{Nonlinear integral equations}
Let us introduce functions $\check{T}_1(\theta)$ and $\check{T}_2(\theta)$ defined by
\begin{equation}
\check{T}_1(\theta):=T_1(z)/\mu^{(1)}(\theta|\{h_j^{(1)}\})\qquad
\check{T}_2(\theta):=T_2(z)/\mu(\theta|\{h_j\}),
\end{equation}
where $\mu^{(1)}(\theta|\{h_j^{(1)}\})$ has simple zeros on the real axis, besides at the origin, at $h^{(1)}_j$ and $\mu(\theta|\{h_j\})$ at $h_j$. 
%
Then $\check{T}_1(\theta)$ and $\check{T}_2(\theta)$ have the property of being analytic and nonzero near the real axis.

\begin{figure}[!h]
\begin{center}
\includegraphics[scale=0.25]{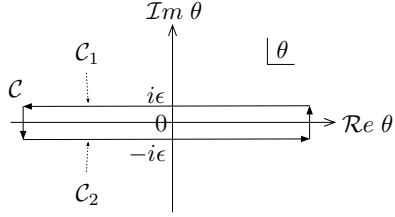}
\caption{The integral contour of the transfer matrices $T_1(\theta)$ and $T_2(\theta)$ taken to enclose all the real roots and holes without any complex roots. 
The small value $\epsilon$ is taken such that $0<|\cI m(\theta_j)|-i\pi/2<\epsilon,\,\forall j$. }\label{fig:contour*}
\end{center}
\end{figure}

Using the Cauchy theorem for the Fourier transform of $\check{T_1}(\theta)$ and $\check{T}_2(\theta)$ by choosing contours enclosing the real axis (Figure \ref{fig:contour*}): 
\begin{equation}
\oint_\cC d\theta\; e^{ik\frac{\pi\theta}{\gamma}} [\ln\check{T_1}(\theta)]''=0
\qquad
\oint_\cC d\theta\; e^{ik\frac{\pi\theta}{\gamma}} [\ln\check{T_2}(\theta)]''=0,
\end{equation}
the nonlinear integral equations are derived in logarithmic forms for $\theta$ in the fundamental analyticity strip $|\cI m\; \theta| < \pi$: 
\begin{equation}\label{NLIE}
\begin{split}
&\ln b(\theta) = -(G*_{\cC_1} \ln B)(\theta) - (G*_{\cC_2} \ln \bar{B})(\theta) 
+ (G_2 ^{[-\frac{\pi}{2}]}*_{\cC_2} \ln Y)(\theta)
\\
&\hspace{16mm}
+ i D_{\rm bulk}(\theta) + i D_{\rm B}(\theta) + i D(\theta) 
+ i \pi C_b
\\
&\ln y(\theta) = -(G_2^{[-\frac{\pi}{2}]}*_{\cC_1} \ln \bar{B})(\theta) + (G_2^{[\frac{\pi}{2}]}*_{\cC_2} \ln B)(\theta)
\\
&\hspace{16mm} 
+ i D_{\rm SB}(\theta) + i \lim_{\eta\rightarrow 0}D_2\left(\theta+\frac{i\pi}{2}-i\eta\right) + i\pi C_y. 
\end{split}
\end{equation}
Here we introduced the following notations; the convolution with integral contour along the real axis is written by
\begin{equation}
(f*g)(\theta) := \int_{-\infty}^{\infty} d\theta' f(\theta - \theta') g(\theta')
\end{equation}
and for the convolution along $\cC$ by 
\begin{equation}
(f*_{\cC}g)(\theta) := \int_{\cC} d\theta' f(\theta - \theta') g(\theta'). 
\end{equation}
For simplicity, we express the function with a shifted variable by
\begin{equation}
f^{[\pm \eta]}(\theta) = f(\theta \pm i\eta). 
\end{equation}
The functions $G(\theta)$ and $G_2(\theta)$ are kernels of the convolutions in the nonlinear integral equations: 
\begin{align}
&G(\theta):=\int_{-\infty}^{\infty}dk\; \frac{e^{-ik \frac{\gamma \theta}{\pi}}}{2\pi}
\frac{\sinh\frac{\pi - 3\gamma}{2}k}{2\sinh\frac{\pi-2\gamma}{2}k \cosh\frac{\gamma}{2}k}
=\frac{1}{2\pi i} \frac{d}{d\theta} \ln S(\theta) 
\label{kernel_G}
\\
&G_2(\theta):=\frac{1}{2\pi}\int_{-\infty}^{\infty}dk\; \frac{e^{-ik \frac{\gamma \theta}{\pi}}}{e^{\frac{\gamma k}{2}} + e^{-\frac{\gamma k}{2}}}
= \frac{1}{2\pi \cosh\theta}, 
\end{align}
from which one may notice that the function $G(\theta)$ is related to the soliton-soliton scattering amplitude (\ref{kernel_G}). 
%
%
%
%

The second and the fourth terms of (\ref{NLIE}) are regarded as \emph{particle source terms} in the sense of the quantum field theory. 
It is useful for writing down the excitation particle terms $\widehat{D}(k)$ and $\widehat{D}_2(k)$ to introduce \emph{effective roots}: 
\begin{equation}
\tilde{\theta}_j := \theta_j - \frac{i\pi}{2} {\rm sign}(\cI m\; \theta_j). 
\end{equation}
Then we have the source terms as follows: 
\begin{align}
&D(\theta) = z_H(\theta)+ z_{H^{(1)}}(\theta) 
- z_S(\theta) 
- z_C(\theta) - z_W(\theta) - z_{SC}(\theta)
\\
&D_2(\theta) = \zeta_H(\theta) 
-  \zeta_S(\theta)
- \zeta_C(\theta) - \zeta_W(\theta) - \zeta_{SC}(\theta), 
\end{align}
where each function is given by 
\begin{equation}
\begin{split}
& z_H(\theta) = \sum_{j=-A}^{A} \chi(\theta-\alpha_j) \hspace{8mm}
z_{H^{(1)}} = \sum_{j=-A}^{A} \chi_2(\theta-\alpha_j) \\
& z_S(\theta) = \sum_{j=-A}^A [\chi(\theta-\alpha_i) + \chi(\theta-\bar{\alpha}_j)] \\
& z_C(\theta) = \sum_{j=-A}^A \chi(\theta-\widetilde{\alpha}_j) \qquad
z_W(\theta) = z_{SC}(\theta) = \sum_{j=-A}^A \chi(\theta-\widetilde{\alpha}_j)_{\rm II}
\end{split}
\end{equation}
and 
\begin{equation}
\begin{split}
& \zeta_H(\theta) = \sum_{j=-A}^A \chi_2(\theta-\alpha_j) \qquad
\zeta_S(\theta) = \sum_{j=-A}^A [\chi_2(\theta-\alpha_j) + \chi_2(\theta-\bar{\alpha}_j)] \\
& \zeta_C(\theta) = \sum_{\j=-A}^A \chi_2(\theta-\widetilde{\alpha}_j) \hspace{8mm}
\zeta_W(\theta) = \zeta_{SC}(\theta) = \sum_{j=-A}^A \chi_2(\theta-\widetilde{\alpha}_j). 
\end{split}
\end{equation}
Here we denoted the number of species by $A$ and rapidity by $\alpha_j$. The notation $\bar{\alpha}$ is used for representing the complex conjugate of $\alpha$. Each function is defined by 
\begin{equation}
\chi'(\theta) = 2\pi G(\theta)
\qquad
\chi'_2(\theta) = 2\pi G_2(\theta)
\end{equation}
and for the second determination of the function $\chi(\theta)$ by 
\begin{equation}
\chi(\theta)_{\rm II} = \chi(\theta) + \chi(\theta - i\pi \; {\rm sign}(\cI m\;\theta)). 
\end{equation}
The bulk term $D_{\rm bulk}(\theta)$ is proportional to the number of sites $N$: 
\begin{equation}
D_{\rm bulk}(\theta) = N \arctan\frac{\sinh\theta}{\cosh\Theta}
\end{equation}
and each of boundary terms $D_{\rm B}(\theta)$ and $D_{\rm SB}(\theta)$ consists of three parts: 
\begin{align}
&D_{\rm B}(\theta) = F(\theta;H_+) + F(\theta;H_-) + J(\theta)
\\
&D_{\rm SB}(\theta) = F_y(\theta;H_+) + F_y(\theta;H_-) + 2 G_2\left(\theta+\frac{i\pi}{2}\right). 
\end{align}
The function $J(\theta)$ is given by 
\begin{equation}
J'(\theta) = \int_{-\infty}^{\infty}dk\; \frac{e^{-ik\frac{\gamma \theta}{\pi}}}{2\pi}
\frac{\cosh\frac{\gamma}{4} k \sinh\frac{\pi - 3\gamma}{4} k}{\cosh\frac{\gamma}{2} k \sinh\frac{\pi - 2\gamma}{4} k}. 
\end{equation}
The functions $F(\theta)$ and $F_2(\theta)$ have different forms depending on the values of boundary parameters: 
\begin{align}\label{boundary_dependent_part}
&F'(\theta; H)=
\begin{cases}
\displaystyle
{\rm sign}(H) \int_{-\infty}^{\infty} dk\; \frac{e^{-ik\frac{\gamma \theta}{\pi}}}{2\pi}
\frac{\sinh\frac{\pi - \gamma |H|}{2} k}{2\cosh\frac{\gamma}{2} k \sinh\frac{\pi-2\gamma}{2} k}
&
|\cI m\;\theta| > \frac{\pi (1 - |H|)}{2}
\vspace{2mm}\\
\displaystyle
\int_{-\infty}^{\infty} dk\; \frac{e^{-ik \frac{\gamma \theta}{\pi}}}{2\pi}
\frac{\sinh\frac{-\pi - \gamma H + 2\gamma}{2} k} {2\cosh\frac{\gamma}{2} k \sinh\frac{\pi-2\gamma}{2} k}
&
|\cI m\; \theta| < \frac{\pi (1- |H|)}{2}
\end{cases}
\\
&
F'_y(\theta;H) =
\begin{cases}
0 & |\cI m\; \theta| > \frac{\pi (1 - |H|)}{2}
\\
G_2\left(\theta + \frac{i \pi H}{2} \right) - G_2\left(\theta - \frac{i\pi H}{2}\right) & |\cI m\; \theta| < \frac{\pi (1- |H|)}{2}. 
\end{cases}
\end{align}
The integration constants $i \pi C_b$ and $i \pi C_y$ are determined from the asymptotic behaviors of the nonlinear integral equations. 

Before ending this subsection, we give a brief comment on the nonlinear integral equations for $\theta$ with the imaginary part outside the fundamental analyticity strip. In this case, the basic structure of the nonlinear integral equations is unchanged by replacing each function by its second determination: 
\begin{equation}
\begin{split}
&\ln b(\theta) = -(G*_{\cC_1} \ln B)(\theta)_{\rm II} - (G*_{\cC_2} \ln \bar{B})(\theta)_{\rm II} 
+ (G_2 ^{[-\frac{\pi}{2}]}*_{\cC_2} \ln Y)(\theta)_{\rm II}
\\
&\hspace{16mm}
+ i D_{\rm bulk}(\theta)_{\rm II} + i D_{\rm B}(\theta)_{\rm II} + i D(\theta)_{\rm II} 
+ i \pi C^{\rm II}_b
\\
&\ln y(\theta) = -(G_2^{[-\frac{\pi}{2}]}*_{\cC_1} \ln \bar{B})(\theta)_{\rm II} + (G_2^{[\frac{\pi}{2}]}*_{\cC_2} \ln B)(\theta)_{\rm II}
\\
&\hspace{16mm} 
+ i D_{\rm SB}(\theta)_{\rm II} + i D_2(\theta)_{\rm II} + i\pi C^{\rm II}_y. 
\end{split}
\end{equation}
The integral constants $i \pi C^{\rm II}_b$ and $i \pi C^{\rm II}_y$ are determined in a similar way to $i \pi C_b$ and $i \pi C_y$ from the asymptotic behaviors, though in general, they have different values for these two cases.

\subsection{Sum rules}
We discussed so far the nonlinear integral equations without referring the restriction of the number of roots and holes. 
Actually, they are not arbitrary but are constrained by the so-called \emph{sum rules}. 
The sum rules are derived from the analysis of asymptotic behaviors of the nonlinear integral equations:  
\begin{align}
&N_H - 2N_S = M_C + 2(S + M_W + M_{SC}) 
\non\\
&\hspace{22mm}
- \left\lfloor \frac{3\gamma(2S+H+1)}{\pi}\right\rfloor
+ \left\lfloor \frac{\gamma(2S+H+1)}{\pi}\right\rfloor
\\
&N_1 - 2N_{S}^{R} = S + M_W + M_{SC} + M_C^{(2)} - M_R  
- \left\lfloor \frac{\gamma (2S+H+1)}{\pi} \right\rfloor, 
\end{align}
where $N_S^R$ is the number of real special objects and $M_C^{(2)}$ is the number of close roots whose imaginary parts lie on $(\pi/2,\pi)$ or $(-\pi,-\pi/2)$. The notation $\lfloor x \rfloor$ was used to express the largest integer part of $x$. 

\subsection{Scaling limit}
The nonlinear integral equations of the lattice SUSY sine-Gordon model are identical to those for the original SUSY sine-Gordon model in the scaling limit. 
The scaling limit is realized by taking the three limits $N, \Theta\rightarrow \infty$ and $a\rightarrow 0$ ($a N:=L$) by keeping the soliton mass $\cM$ finite. 
These quantities appear only through the bulk term (\ref{NLIE}) which results in 
\begin{equation}
N \arctan\frac{\sinh \theta}{\cosh \Theta} 
\underset{\text{scaling limit}}{\longrightarrow}
2i \cM L \sinh\theta. 
\end{equation}

\section{Boundary bound states and the ground state of BSSG model}\label{sec:bbs}
In the context of the $S$-matrix theory, rapidity of particles bounded at the boundaries emerge as poles in the reflection matrix. On the other hand, boundary bound states are obtained as pure imaginary roots from the Bethe ansatz context. 
These two objects nevertheless does not coincides in the finite system, due to the finite-size deviation of the Bethe roots. They have the same values only in the infinite-volume limit $\cM L \rightarrow \infty$, which we are going to consider in this section.

\subsection{Large volume limit}
In the relation with the quantum field theory, we call the $\cM L\rightarrow \infty$ limit the infrared (IR) limit. The nonlinear integral equations (\ref{NLIE}) are simplified in the IR limit into
\begin{equation}\label{bare_NLIE}
\begin{split}
&\ln b(\theta) = (G_2^{[-\frac{\pi}{2}]}*_{\cC_2} \ln Y) (\theta) + 2i\cM L \sinh\theta + i D_{\rm B}(\theta) + i D(\theta) + i\pi C_b
\\
&\ln y(\theta) = i D_{\rm SB} (\theta) + i D_2(\theta) + i\pi C_y. 
\end{split}
\end{equation}
One may notice that we also took the scaling limit.

Suppose the Dirac sea consists of the pure two-strings without particles for $H_{\pm} > 1$. The one-soliton excitation is then described by appearance of a hole without any other objects; 
Then the nonlinear integral equations with a hole are equations for the behavior of one soliton in a finite interval $L$. 
Therefore the sub-leading terms of the nonlinear integral equations (\ref{bare_NLIE}) give the scattering amplitude on the ground state boundary: 
\begin{equation}
\exp[(G_2^{{[-\frac{\pi}{2}]}}*_{\cC_2} \ln Y) (\theta) + i D_{\rm B}(\theta)]\Big|_{H_->1}
=
-\frac{\cR_{\rm SSG}^{|0\rangle \otimes |\frac{1}{2}\rangle} (\theta;\lambda,\xi_-) 
\cR_{\rm SSG}^{|0\rangle \otimes |\frac{1}{2}\rangle} (\theta;\lambda,\xi_+)}
{S(2\theta)}, 
\end{equation}
from which one can read off the relations between boundary parameters of the $S$-matrix theory and the lattice-regularized model: 
\begin{equation}\label{identification}
\begin{split}
&t - H = 1 + \frac{2 \xi}{\pi \lambda}
\\
&t - 2 = \frac{1}{\lambda}. 
\end{split}
\end{equation}
Here we set a superscript $|B\rangle \otimes |a\rangle$ to $\cR_{\rm SSG}$, whose first component indicates the sine-Gordon part of a boundary state and the second the RSOS part. There are no solitons as the state $|0\rangle$ represents with the RSOS index $\frac{1}{2}$ due to non-degeneracy of the ground state \cite{bib:BPT02}. We will denote the existence of a pure imaginary root $\theta = i\nu_0$ by a superscript $|\nu_0\rangle$ and the corresponding RSOS index by one of the doublet RSOS states $|a\rangle$ ($a=0,1$). 

From now on, we call the nonlinear integral equations without the source terms as the bare nonlinear integral equations. Once again the bare nonlinear integral equations given in (\ref{bare_NLIE}) were written for the pure two-string ground state. On the other hand, the parameter relations (\ref{identification}) indicate that the first pole in the reflection matrix $\theta =i\nu_0$ gets into the physical strip for $0<H_-<1$, that is, a boundary bound state emerges in this regime. 

Now the question is which state the bare nonlinear integral equations are describing. For simplicity, we let only $H_-$ move by fixing $H_+$ at a value larger than $1$. Thus, boundary bound states would appear only at the $(-)$-boundary. Let us first sum up three different forms of the boundary term $D_{\rm B}(\theta)$ obtained by analytic continuation with respect to $H_-$ (Table \ref{boundary_regimes}). 
\begin{table}[!h]
\begin{center}
\begin{tabular}{llll}
\hline
\hline
(a) & $H>1$ & $D_{\rm B}(\theta)|_{H>1}$ & $D_{\rm SB}(\theta)|_{H>1}$
\\
(b) & $1>H>-1$ & $D_{\rm B}(\theta)|_{1>H>-1}$ & $D_{\rm SB}(\theta)|_{1>H}$
\\
(c) & $-1>H$ & $D_{\rm B}(\theta)|_{-1>H}$ & $D_{\rm SB}(\theta)|_{1>H}$
\\
\hline
\hline
\end{tabular}
\caption{Three regimes with different forms of boundary terms. }\label{boundary_regimes}
\end{center}
\end{table}
In the case of (a), we obtained (\ref{bare_NLIE}) describes the pure two-string ground state. 
In the regime (b), 
the analytic continuation with respect to $H_-$ makes $D_{\rm B}(\theta)$ into a different form. 
Recall the boundary bootstrap equation for the sine-Gordon part:
\begin{equation}
\cR_{\rm SG}^{|\nu_0\rangle}(\theta;\lambda,\xi) = \cR_{\rm SG}^{|0\rangle}(\theta;\lambda,\xi) S(\theta -i\nu_0) S(\theta +i\nu_0), 
\end{equation}
which precisely gives the expression of $D_{\rm B}(\theta)$ for the regime (b). 
This means the analytic continuation of the nonlinear integral equations naturally involves the appearance of boundary excitation as the appearance of a hole at $\theta = i\nu_0$. 

Thus the straightforward computation leads us to obtain that the reflection amplitude on the first excited boundary, given by the appearance of pure imaginary holes at $\theta=\pm \pi(1-H_-)/2$, coincides with what is described by the bare nonlinear integral equations for $1>H_->-1$:
\begin{equation}
\exp[(G_2^{{[-\frac{\pi}{2}]}}*_{\cC_2} \ln Y) (\theta) + i D_{\rm B}(\theta)]\Big|_{1>H_->-1}
=
-\frac{\cR_{\rm SSG}^{|\nu_0\rangle \otimes |a\rangle} (\theta;\lambda,\xi_-) 
\cR_{\rm SSG}^{|0\rangle \otimes |\frac{1}{2}\rangle} (\theta;\lambda,\xi_+)}
{S(2\theta)}. 
\end{equation}

On the other hand, the bare nonlinear integral equations for $H_-<-1$ are related to those for $H_->1$ by 
\begin{equation}
\begin{split}
&(G_2^{{[-\frac{\pi}{2}]}}*_{\cC_2} \ln Y) (\theta) + D_{\rm B}(\theta)\Big|_{-1>H_-}\\
&=
(G_2^{{[-\frac{\pi}{2}]}}*_{\cC_2} \ln Y) (\theta) + D_{\rm B}(\theta)\Big|_{H_->1}
+ 
\chi_2\left(\theta-\frac{i\pi H_-}{2}\right) + \chi_2\left(\theta+\frac{i\pi H_-}{2}\right), 
\end{split}
\end{equation}
which implies the appearance of type-1 holes at $\theta=- i\pi H_-/2$. As was already referred, a SUSY sine-Gordon particle is interpreted as a hole in the $2$-strings. Since the type-1 hole lies in the distribution of real roots, this should be considered as an excitation for the system with the ground state described by the pure two-strings. 

We give an interpretation on this situation; the pure two-string ground state or states consisting of ordinary holes are realized by an even number of Bethe roots. If the system is encoded by an odd number of Bethe roots from the beginning, not every root can make pairs and thus strong enough boundary field arrests only one root. Thus, our suggestion is that the bare nonlinear integral equations for $H_->1$ and $H_-<-1$ describe difference sectors of the SUSY sine-Gordon model, which cannot be mixed simply by parameter transformations or particle excitations.\footnote{The bare nonlinear integral equations for $H_->1$ cannot reproduce all the eigenvalues of the RSOS part of the scattering matrix \cite{bib:HRS06}. We expect the two of them which did not show up would be obtained from those for $H<-1$. }

\subsection{Ground state}
In this section, we discuss the ground state of the SUSY sine-Gordon model with Dirichlet boundary conditions on a finite interval. The similar analysis has done for the spin-$1/2$ chain under existence of boundary magnetic fields \cite{bib:SW09} and they found 10 regimes with different structures for the ground state. 

The main protagonists of the analysis are pure imaginary roots/holes and poles in the auxiliary functions. As is referred in the previous section, the poles which may appear for a given value of $H_{\pm}$ are naturally taken into consideration by analytic continuation of the nonlinear integral equations. 
 
\begin{figure}[!h]
\begin{center}
\subfigure[(left) Zeros of $1+a_1(\theta)$; (right) $B(\theta)$ for $H_+=1.5$ and $H_-=2.2$.]
{\includegraphics[scale=0.5]{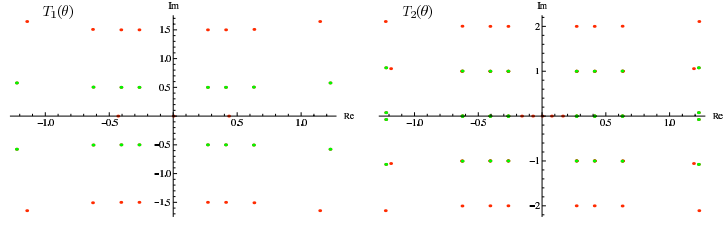}}
\label{fig:zeros-a}
\subfigure[(left) Zeros of $1+a_1(\theta)$; (right) $B(\theta)$ for $H_+=1.5$ and $H_-=0.3$.]
{\includegraphics[scale=0.5]{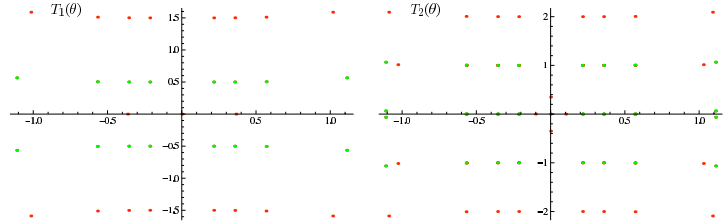}}
\label{fig:zeros-b}
\subfigure[(left) Zeros of $1+a_1(\theta)$; (right) $B(\theta)$ for $H_+=1.5$ and $H_-=-0.5$.]
{\includegraphics[scale=0.5]{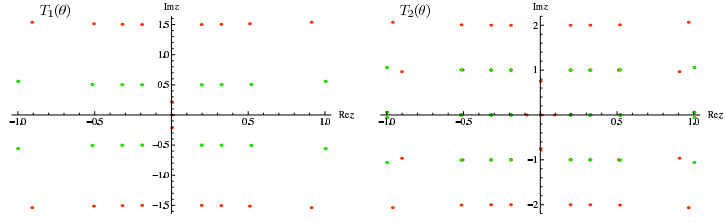}}
\label{fig:zeros-c}
\subfigure[(left) Zeros of $1+a_1(\theta)$; (right) $B(\theta)$ for $H_+=1.5$ and $H_-=-1.8$.]
{\includegraphics[scale=0.5]{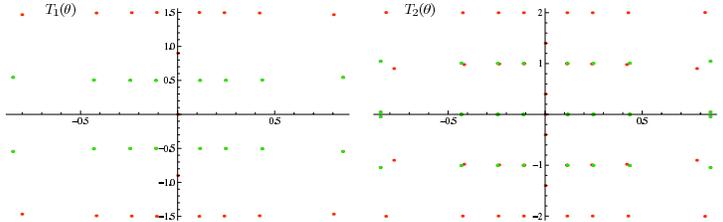}}
\label{fig:zeros-d}
\end{center}
\caption{Zeros of the auxiliary functions $1+a_1(\theta)$ and $B(\theta)$ is plotted for $N=8$ with $8$ roots in the homogeneous and isotropic limit. Roots are plotted by green dots and hole-type zeros by red dots.}\label{fig:zeros}
\end{figure}

The degree analysis indicates that being set as $w=e^{\theta}$, there exist $N+1$ holes in $1+a_1(\theta)$ and $N+2$ holes in $B(\theta)$ for the system of length $N$ under existence of $N$ roots, as is also numerically checked in Figure \ref{fig:zeros}.
Now we look into how these holes move depending on the values of boundary parameters. 
Following the previous section, we let only one of the boundary parameters $H_-$ move by fixing the other $H_+$ at arbitrary value larger than $1$. 
For boundary field $H_-$ larger than $1$, all holes lie on the real axis (Figure \ref{fig:zeros-a}). As boundary field strengthens, that is, as the value of $H_-$ becomes smaller, one of the holes in $T_2(\theta)$ moves toward the origin and then on the imaginary axis after $H_-$ passes through $1$ from above (Figure \ref{fig:zeros-b}). This hole seems to be fixed at $\theta = i\pi (1-H_-)/2$ up to exponentially small correction of the system size. 
At the same time, one of the poles in the auxiliary function $B(\theta)$ locating at $\theta = -i\pi (1-H_-)/2$ also passes the origin and subsequently the Bethe equations admit a pure imaginary root. The appearance of a pure imaginary root is understood in the context of the Bethe ansatz equations for the large enough system as the pole with a positive value of the imaginary part is compensated by the term of the $N$th power which vanishes as $N$ gets large. As a result, the Bethe ansatz equations with large enough $N$ admit pure imaginary roots for $H_-<1$. 
Here we remark that a hole lies on the distribution of the two-strings. Thus, it should considered that two roots are lost when a hole emerges. Assuming the value of hole rapidity is given by string-ceter rapidity of two roots, we conclude that two imaginary roots appear at $\theta=-i\pi H_-/2$ and $-i\pi(H_--2)/2$.  
One can easily check that the wave function composed by imaginary rapidity exponentially small from the boundary. For this reason, the imaginary roots can be understood to form boundary bound states. 

Now we consider smaller $H_-$ than $1$. When $H_-$ crosses $0$, one of the holes in $T_1(\theta)$ reaches the origin moving on the real axis. This hole then moves up along the imaginary axis as $H_-$ decreases fixed at $\theta = -i\pi H_-/2$ up to exponentially small finite-size correction (Figure \ref{fig:zeros-c}). 
The auxiliary function $1+a_1(\theta)$ has a pole of the same value as the hole-type zero of $T_1(\theta)$ up to the correction and picks it up to the imaginary axis. Thus a pure imaginary root $\theta = -i\pi H/2$ appears. 
One has to be careful that this time the hole emerges from $1+a_1(\theta)$, which counts real roots. Thus, the appearance of a type-1 hole is achieved simply by removing one root. 
Therefore, a type-1 hole generates a different boundary bound state from the two-imaginary-root case. 

Finally when $H_-$ becomes smaller than $-1$, the other hole of $T_2(\theta)$ reaches the origin and goes up on the imaginary axis (Figure \ref{fig:zeros-d}). 
Being fixed at $\theta=-i\pi(H_-+1)$, this hole is picked up by a pole in the auxiliary function $B(\theta)$. Thus another imaginary roots appear at $\theta = -i\pi H_-/2$ and $-i\pi(H_-+2)/2$ in this regime.


\subsection{Symmetry with respect to boundary parameters}
We give some comments on the symmetries of the SUSY sine-Gordon model. 
From the action of the model (\ref{BSSG_action}), we obtain that the theory is invariant under the transformations: 
\begin{align}
&\varphi_{\pm} \rightarrow \frac{2\pi}{\beta} -\varphi_{\pm}
\\
&\varphi_{\pm} \rightarrow \frac{2\pi}{\beta} + \varphi_{\pm}
\\
&\varphi_{\pm} \rightarrow -\varphi_{\pm}. 
\end{align}
In comparison with the isotropic limit of the model, it is useful to rewrite the symmetry in terms of the lattice boundary parameter $H_{\pm}$ by utilizing the relations (\ref{identification}): 
\begin{enumerate}
\item[(a)]	$\varphi_{\pm}\;\rightarrow\;2\pi/\beta-\varphi_{\pm}$
\qquad
$\xi_{\pm}\;\rightarrow\;2\xi_0-\xi_{\pm}$
\qquad 
$H_{\pm}\;\rightarrow\;-H_{\pm} - 2$
\item[(b)]	$\varphi_{\pm}\;\rightarrow\;2\pi/\beta+\varphi_{\pm}$
\qquad
$\xi_{\pm}\;\rightarrow\;2\xi_0+\xi_{\pm}$
\qquad 
$H_{\pm}\;\rightarrow\;H_{\pm} - 2t$
\item[(c)]	$\varphi_{\pm}\;\rightarrow\;-\varphi_{\pm}$
\hspace{17mm}
$\xi_{\pm}\;\rightarrow\;-\xi_{\pm} $
\hspace{13mm} 
$H_{\pm}\;\rightarrow\;-H_{\pm}+2(t-1)$
\end{enumerate}
where $2\xi_0:=\pi(2\lambda+1)$ and $\xi_{\pm}=2\pi\varphi_{\pm}/\beta$. 

\begin{figure}[!h]
\begin{center}
\includegraphics[scale=0.75]{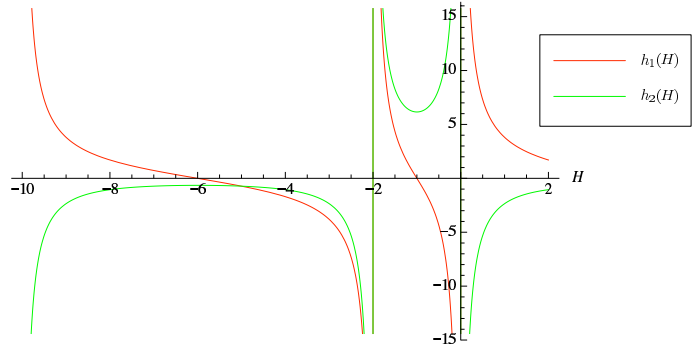}
\caption{Energy of boundary terms in the spin-$1$ XXZ model with boundary magnetic fields. }\label{fig:benergy}
\end{center}
\end{figure}

The corresponding spin system to the SUSY sine-Gordon model has the following interactions; 
\begin{equation}\label{hamiltonian}
\cH=\sum_{j=1}^{N-1}H_{j,j+1} + \cH_{\rm B}. 
\end{equation}
The bulk term is given by 
\begin{equation}
\begin{split}
H_{j,j+1} =& T_j-(T_j)^2-2(\sin\gamma)^2\;[T_j^z+(S_j^z)^2+(S_{j+1}^z)^2-(T_j^z)^2]\\
&+4\left(\sin\frac{\gamma}{2}\right)^2\;(T_j^{\bot}T_j^z+T_j^zT_j^{\bot}),
\end{split}
\end{equation}
where 
\begin{equation}
T_j=\vec{S}_j\cdot \vec{S}_{j+1}\qquad
T^{\bot}_j=S^x_jS^x_{j+1}+S^y_jS^y_{j+1}\qquad
T^z_j=S^z_jS^z_{j+1}
\end{equation}
and the boundary term by
\begin{equation}
\cH_{\rm B} = h_1(H_-) S_1^z + h_2(H_-)(S_1^z)^2
+ h_1(H_+) S_N^z +  h_2(H_+) (S_N^z)^2,  
\end{equation}
where
\begin{align}
&h_1(H) = \frac{1}{2} \sin(2\gamma) \left[\cot\left(\frac{\gamma H}{2}\right) + \cot\left(\frac{\gamma H}{2}+\gamma\right)\right]
\\
&h_2(H) = \frac{1}{2} \sin(2\gamma) \left[-\cot\left(\frac{\gamma H}{2}\right) + \cot\left(\frac{\gamma H}{2}+\gamma\right)\right]. 
\end{align}
In Figure \ref{fig:benergy}, the boundary energies of the spin chain are depicted.   
One obtains the same symmetries of the SUSY sine-Gordon model in the boundary energies of the spin chain. Thus, the three symmetries of the quantum field theory survive even after taking the isotropic limit, which supports validity of our numerical results in the analysis of the SUSY sine-Gordon model. 

%
%

\section{Concluding remarks}
In this paper, we studied the ground state of the SUSY sine-Gordon model. The structure of the ground state roots strongly depends on the values of boundary parameters. What was found here is that there are four regimes which are characterized by different features of Bethe roots. 
The detailed structure is; 
\begin{enumerate}\setcounter{enumi}{1}
	\item in the regime either $H_->1$ or $H_-<-2$, the ground state is described by the pure two-strings. 
	\item For $1>H_->0$, the ground state include one soliton with rapidity $\theta = i\pi(1-H_-)/2$. In the context of the Bethe ansatz equations, the ground state is encoded by the two-strings and two imaginary roots $\theta = -i\pi H_-/2$ and $-i\pi(H_--2)/2$. 
	\item In the regime $0>H_->-1$, a type-1 hole contributes to the ground state. Rapidity of this hole is given by the corresponding Bethe root itself $\theta = -i\pi H_-/2$. This situation is realized only when the original system is characterized by the odd number of roots. Thus, the lowest energy state in this regime belongs to a different class from the other regimes. 
	\item Finally for $-1>H_->-2$, the ground state is given again by the two-strings and one soliton with pure imaginary rapidity, but at $\theta = -i\pi(H_-+1)/2$ this time. Correspondingly, the ground state includes two pure imaginary roots at $\theta = -i\pi H_-/2$ and $-i\pi(H_-+2)/2$. 
\end{enumerate}

We also discussed what is described by the analytic continuation of the nonlinear integral equations with respect to the boundary parameters in the Fourier space. It was found that 
\begin{enumerate}
	\item the bare nonlinear integral equations for $H_->1$ describes the two-string ground state. 
	\item In the regime $1>H_->-1$, they are written for the state which includes a hole at $\theta = i\pi(1-H_-)/2$. 
	\item the bare nonlinear integral equations for $H_-<-1$ are the equations for the state with a type-1 hole at $\theta = -i\pi H_-/2$, which belongs to the different class of the above two. 
\end{enumerate}
%
%

The nonlinear integral equations provide the sum rules, the restriction on the number of various roots and holes, from their asymptotic behavior. The sum rules help us to know the possible configurations of roots and holes. 
We expect that the all particle and reflection amplitudes of the SUSY sine-Gordon model would be recasted from configurations of holes and roots, which include close roots or wide roots.
It is also an interesting open problem to study scatterings on the system whose ground state is  encoded by the odd number of roots, which may give the lacked two eigenvalues of the RSOS part of the reflection matrix. 

We briefly remarked the symmetry of the SUSY sine-Gordon model with respect to the boundary parameters. They are translated into the symmetries of lattice boundary parameters via the identification obtained from the comparison of the sub-leading term of the nonlinear integral equations with the reflection amplitudes. 
As it was found that these symmetries survive in the homogeneous limit from the energy plots of the boundary terms, we expect the ground state of the spin chain has the same structure as the SUSY sine-Gordon model.

\section*{Acknowledgements}
The author would like to be gratitude to F. G\"{o}hmann, A. Kl\"{u}mper, S. Miyashita, and J. Suzuki for helpful discussions and comments. We acknowledge JSPS Research Fellowship for Young Scientists for supporting the beginning of this work. 

This research is supported by the Aihara Innovative Mathematical Modelling Project, the Japan Society for the Promotion of Science (JSPS) through the ``Funding Program for World-Leading Innovative R\&D on Science and Technology (FIRST Program)," initiated by the Council for Science and Technology Policy (CSTP).


\end{document}